\documentstyle[aps]{revtex}
\begin{document}
\twocolumn[\hsize\textwidth\columnwidth\hsize\csname@twocolumnfalse%
\endcsname
\draft

\title{
Asymmetric double barrier resonant tunnelling structures \\
with improved characteristics
	}

\author{Jun-jie Shi\thanks{On leave from Department of Physics, 
        Henan Normal University, Xinxiang 453002, Henan, People's Republic of 
        China.}$^{1,3}$, Barry C.\ Sanders$^1$ and Shao-hua Pan$^{1,2,3}$\\
	$^1$ Department of Physics, Macquarie University,
	Sydney, New South Wales 2109, Australia \\
	$^2$ Institute of Physics, Chinese Academy of Sciences,
	P.\ O.\ Box 603, Beijing 100080, P.\ R.\ China	\\
	$^3$ China Center of Advanced Science and Technology 
	(World Laboratory),
	P.\ O.\ Box 8730, Beijing 100080, P.\ R.\ China
}
\date{\today}
\maketitle

\begin{abstract}

	We present a self-consistent calculation, based on the global coherent tunnelling model, and  show that structural asymmetry of double barrier resonant tunnelling structures  significantly modifies the current-voltage characteristics compared to the symmetric structures. In particular, a suitably designed asymmetric structure can produce
much larger peak current and absolute value of the negative differential conductivity than its commonly used symmetric counterpart.

\end{abstract}



]

\narrowtext

Significant progress in the growth of semiconductor materials has 
re-invigorated research in double-barrier resonant tunnelling structures 
(DBRTSs) and their applications\cite{Sun98}. The key feature of the 
DBRTS\cite{Tsu73} which promises benefits is the negative differential 
resistance (NDR)\cite{Sollner83}. For example, the NDR can be exploited in 
high-speed analog-to-digital converters\cite{Sen87} and parity  generators\cite{Capasso89}. We propose asymmetric structure design to improve the DBRTS characteristics, in particular the peak current and the NDR. We show that an appropriate asymmetric design can yield improved characteristics. Our simulation employs the self-consistent global coherent tunnelling model which is valid at and near the resonance bias. 

In the global coherent tunnelling model we can express the 
three-dimensional electronic wave function for a DBRTS as a product of the bulk Bloch wave function and a one-dimensional envelope wave function $\psi_{E_z}(z)$
along the growth axis $(z)$ of the DBRTS, 
satisfying the Schr\"odinger equation
\begin{eqnarray}
&&-{\hbar^2\over 2}{d\over dz}\left[{1\over m(z)}{d\over dz}
	\psi_{E_z}(z)\right]
	+\left[V(z)-e\Phi(z)\right]\psi_{E_z}(z) \nonumber \\ &&
=E_z\psi_{E_z}(z), 
\end{eqnarray}
with $m(z)$ the electron effective mass, $-e$ the electron charge, $V(z)$ 
the conduction-band offset, and $\Phi(z)$  the Hartree potential satisfying 
the 
Poisson equation
\begin{equation}
{d\over dz}\left[\epsilon(z){d\Phi(z)\over dz}\right]=
-e\left[N_D^+(z)-n(z)\right], 
\end{equation}
where $\epsilon(z)$  is the dielectric constant, $N_D^+(z)$ is the density 
of ionised donors, and $n(z)$ is the electron density which can be expressed as
\begin{equation}
n(z)=n^{(E\rightarrow C)}(z)+n^{(C\rightarrow E)}(z). 
\end{equation}
Here $n^{(E\rightarrow C)}(z)$ ($n^{(C\rightarrow E)}(z)$)
   is the density of electrons tunnelling from the emitter to collector (from 
the collector to emitter). In the following we write explicitly only for
$n^{(E\rightarrow C)}(z)$, and $n^{(C\rightarrow E)}(z)$  can be treated 
similarly.
\begin{eqnarray}
n^{(E\rightarrow C)}(z)&=&\frac{L_Ek_BT}{\pi^2\hbar^3}
\left(\frac{m_E}{2}\right)^{3/2} \nonumber \\ && \times
\int_0^\infty \frac{ {\rm ln}\{1+{\rm exp}[(E_F-E_z)/k_BT]\} }{\sqrt{E_z}}
	\nonumber \\ && \times 
\left|\frac{1}{\sqrt{B}}\psi_{E_z}^{(E\rightarrow C)}(z)\right|^2dE_z,
\end{eqnarray}
where $L_E$  is the emitter length, $k_B$ and $T$ are, respectively, the 
Boltzmann constant and temperature, $E_F$  is the Fermi energy level, and 
$m_E$  is the electron effective mass in the emitter. In our iterative 
numerical calculation for self-consistent solutions of the coupled 
Schr\"odinger and Poisson equations, the normalization constant $B$ is 
determined by
\begin{eqnarray}
B &=& \frac{k_BT}{\pi^2\hbar^3n^{(E)}}\left(\frac{m_E}{2}\right)^{3/2}
	\nonumber \\ && \times
\int_0^\infty \frac{ {\rm ln} \{1+{\rm exp}[(E_F-E_z)/k_BT]\} }{\sqrt{E_z}}dE_z
	\nonumber \\ && \times
\int_0^L\left|\psi^{(E\rightarrow C)}_{E_z}(z)\right|^2dz, 
\end{eqnarray}
where $n^{(E)}$ is the dopant density in the emitter and~$L$  is the entire 
length of the DBRTS including the emitter and collector.
Equation (5) is derived from the requirement of electric charge neutrality, namely  
$\int_0^L n^{(E\rightarrow C)}(z)dz=L_En^{(E)}$.
A similar treatment for self-consistent solutions has been used by 
Zimmermann et al.\cite{Zimmermann88}. Our numerical calculations show that 
such a normalisation method ensures fast convergence of the solution. 
It is worth mentioning that the appearance of $L_E$ and $n^{(E)}$ on Eqs.~(4) and (5) implies that the density of tunneling electrons depends on the parameters of electron reservoir. In our numerical calculations, a 
standard iteration algorithm has been used: Eq.~(1) is solved with a 
transfer matrix method\cite{Shi98}, and Eq.~(2) is solved in the subsequent 
iterations with Dirichlet boundary conditions, which keep the values of the 
potential at the two boundaries fixed.

We present our numerical results for the following DBRTSs:
$ \mathcal{S}$$(L_E,d_w,d_r;x)$=$n^+$GaAs($L_E$)/GaAs(4{\rm nm})- 
/Al$_x$Ga$_{1-x}$As(3{\rm nm})/GaAs($d_w$)/Al$_{0.3}$Ga$_{0.7}$As($d_r$)/-
GaAs(4{\rm nm})/$n^+$GaAs($L_E$).
The emitter and collector are assumed to be of equal lengths, and the doping 
concentration is assumed to be $n^{(E,C)}=10^{18}$ cm$^{-3}$. Calculations are all performed at $T=300$ K. The electron effective mass in Al$_x$Ga$_{1-x}$As  
is $m=(0.0665+0.0835x)m_0$  with $m_0$  the free electron mass. The conduction-band discontinuity between the GaAs and Al$_x$Ga$_{1-x}$As  is given by
$V(x)=0.6(1266x+260x^2)$ meV. The relative dielectric constant in
Al$_x$Ga$_{1-x}$As is $13.18-3.12x$.
	
Figure~\ref{fig1} presents current-voltage curves for DBRTSs $S(10{\rm nm},5{\rm nm},d_r;0.3)$  with 
the right-barrier thickness  $d_r=2$, 3 and 4 {\rm nm}. The figure shows that the asymmetric structure with a narrower
right barrier ($d_r=2$ {\rm nm}) has a higher peak current ($I_p$) than the symmetric one 
($d_r=3$ {\rm nm}). The trend of increasing $I_p$  with decreasing $d_r$, shown in 
the figure, is consistent with recent experimental results\cite{Schmidt96}. 
This is because the peak transmission coefficient of the electron is higher 
for a structure with narrower right barrier. Moreover, Fig.~\ref{fig1} shows 
that the peak position moves towards higher bias voltage for a DBRTS with 
thicker right barrier. This trend is also consistent with previous 
experimental results\cite{Schmidt96}. However, due to neglect of phase-coherence breaking  
scattering, the present simple model cannot explain the decreasing 
peak-to-valley ratio (PVR) with increasing right barrier width, which was 
observed by previous authors and was attributed to the increase of the valley current and change of PVR caused by electron-electron scattering\cite{Schmidt96}.
	
Figure~\ref{fig2} presents calculated results for DBRTSs  
$S(10{\rm nm},5{\rm nm},3{\rm nm};x)$ with the 
left-barrier Al mole $x=0.2$, 0.3 and 0.4. The figure shows that the peak current and absolute value of the negative differential conductivity are higher for a structure with 
lower $x$ and hence lower left barrier. Although we have not found previously 
published experimental data to test the above calculated results, we can 
attribute the above trend to two physical reasons. First, a structure with a 
lower left barrier has a higher peak transmission coefficient. Secondly, the 
bias raises the left barrier top to the right one. A structure with a lower 
left barrier can compensate the latter effect and hence can cause the left 
and right barrier tops to locate at similar energy levels under resonant bias.
For example, the left and right barrier tops in the $x=0.2$ 
asymmetric DBRTS under resonance bias are closer than that in the 
symmetric one with $x=0.3$. As is known, a DBRTS with the same or similar 
left and right barrier potential can achieve enhanced transmission 
coefficient. Based on the above two physical reasons, we can understand why 
the peak currents and absolute value of the negative differential conductivities
of the structures with $x=0.2$, 0.3 and 0.4 have such a large 
difference as shown in Fig.~\ref{fig2}. In particular, the peak current 
and absolute value of the negative differential conductivity for 
the asymmetric structure with $x=0.2$ are much larger than that of the 
symmetric one ($x=0.3$). Such asymmetric structural effect, which is favorable to 
some device applications of DBRTSs, await experimental confirmation.  

Incidentally, in our calculations shown in Figs.~\ref{fig1} and 2 we 
have set the emitter length $L_E=10$ {\rm nm}. This scale is somewhat shorter than 
that in commonly-used DBRTSs. However, some authors prefer a shorter emitter, 
e.g., $L_E=5$ {\rm nm} in Ref.\cite{Lake92} and in Fig. 2.7 of Ref.\cite{Mizuta95}.

	Besides the above study on the role of asymmetry between the left and 
right barriers, we have also investigated the size effect of the quantum well 
width $d_w$. The calculations show that the peak current increases and the 
peak position shifts to a higher bias voltage if the well width decreases, 
which is in agreement with the previous experiments\cite{Teitsworth94}. As 
for the size effect of the emitter (collector) length $L_E$, our calculations 
show that when the emitter length $L_E$ decreases, the electron accumulation 
in the quantum well becomes smaller which gives rise to a weaker 
self-consistent field resulting in a smaller band bending, and hence the peak 
current increases with decreasing the emitter (collector) length $L_E$. The 
reduction of peak current with increasing emitter length is obvious for
$L_E\leq 10$ {\rm nm} and is shown in Fig.~\ref{fig3}, while it is less obvious for larger $L_E$ (say $L_E>20$ {\rm nm}) and is not shown here. 

	In summary, our self-consistent calculations show that the resonant 
tunnelling current depends sensitively on and hence can be effectively 
controlled by structure parameters, such as the barrier heigths and widths of 
the left and right barriers, and the well width. An asymmetric DBRTS with a 
suitably designed structure has a much larger peak current and absolute value 
of the negative differential conductivity
than the commonly used symmetric one. Moreover, the size effect of the emitter (collector) length is theoretically studied and explained for the first time.

\newpage

\begin{figure}
\caption{
         Current-voltage curves for DBRTSs 
$n^+$GaAs($10{\rm nm}$)/GaAs(4{\rm nm})/Al$_{0.3}$Ga$_{0.7}$As(3{\rm nm})/\\
GaAs($5{\rm nm}$)/Al$_{0.3}$Ga$_{0.7}$As($d_r$)/GaAs(4{\rm nm})/$n^+$GaAs($10{\rm nm}$).  
The solid, dash and dash-dot lines are for  $d_r=2$, 3 and 4 {\rm nm}, respectively.
}
\label{fig1}
\end{figure}

\begin{figure}
\caption{
        Current-voltage curves for DBRTSs  
$n^+$GaAs($10{\rm nm}$)/GaAs(4{\rm nm})/Al$_x$Ga$_{1-x}$As(3{\rm nm})/\\
GaAs(5{\rm nm})/Al$_{0.3}$Ga$_{0.7}$As(3{\rm nm})/GaAs(4{\rm nm})/$n^+$GaAs(10{\rm nm}).
The solid, dash and dash-dot lines are for $x=0.2$, 0.3 and 0.4, respectively.
}
\label{fig2}
\end{figure}

\begin{figure}
\caption{
Current-voltage curves for DBRTSs  
$n^+$GaAs($L_E$)/GaAs(4{\rm nm})/Al$_{0.3}$Ga$_{0.7}$As(3{\rm nm})/\\
GaAs(5{\rm nm})/Al$_{0.3}$Ga$_{0.7}$As(3{\rm nm})/GaAs(4{\rm nm})/$n^+$GaAs($L_E$).
The solid, dash  and dash-dot lines are for  $L_E=5$, 8 and 10 {\rm nm}, respectively.
}
\label{fig3}
\end{figure}

\begin{references} 
\bibitem {Sun98} Sun, J.P., Haddad, G.I., Mazumder, P. and Schulman, J.N., {\em Proc. IEEE}. {\bf 86}, 1998, 641.
\bibitem {Tsu73} Tsu, R. and Esaki, L., {\em \apl} {\bf 22}, 1973, 562; Chang, L.L., Esaki, L. and Tsu, R., {\em \apl} {\bf 24}, 1974, 593.
\bibitem {Sollner83} Sollner, T.C.L.G., Goodhue, W.D., Tannenwald, P.E., 
Parker, C.D. and Peck, D.D., {\em \apl} {\bf 43}, 1983, 588.
\bibitem  {Sen87} Sen, S., Capasso, F., Cho, A.Y., and Sivco, D., {\em IEEE Tran. Electron Devices}, {\bf 34}, 1987, 2185.
\bibitem {Capasso89} Capasso, F., Sen, S., Beltram, F., Lunardi, L.M., Vengurlekar, A.S., Smith, P.R., Shah, N.J., Malik, R.J., and Cho, A.Y., {\em IEEE Tran. Electron Devices}, {\bf 36}, 1989, 2065.
\bibitem {Zimmermann88} Zimmermann, B., Marclay, E., llegems, M. and Gueret, P., {\em J. Appl. Phys}. {\bf 64}, 1988, 3581.
\bibitem {Shi98} Shi, J.-J., Sanders, B.C. and Pan, S.-H., {\em Euro. Phys. Journal B} {\bf 4}, 1998, 113.
\bibitem {Schmidt96} Schmidt, T., Tewordt, M., Haug, R.J., Klitzing, K.V., Sch\"onherr, B., Grambow, P., F\"orster, A. and L\"uth, H., {\em  \apl} {\bf 68}, 1996, 838.
\bibitem {Lake92} Lake, R. and Datta, S., {\em  \prb} {\bf 45}, 1992, 6670.
\bibitem {Mizuta95} Mizuta, H. and Tanoue, T.,  {\em The Physics and Applications of Resonant Tunnelling Diodes}, Cambridge University Press, Cambridge, 1995, p. 23.
\bibitem {Teitsworth94} Teitsworth, S.W., Turley, P.J., Wallis, C.R., Li, W. and Bhattacharya, P.K., {\em Semicond. Sci. Technol}. {\bf 9}, 1994, 508.
\end{references}
\end{document}